\documentclass[12pt]{article}
\usepackage{amsmath}
\usepackage{url}
\usepackage[pdftex]{graphicx}
\usepackage[pagewise]{lineno}

\title{The Web of Life}

\author{Miguel A. Fortuna\footnote{To whom correspondence should be addressed. e-mail: fortuna@ebd.csic.es} , Ra\'ul Ortega, and Jordi Bascompte}
\date
 {
  \vspace{1cm}
  Integrative Ecology Group\\
  Estaci\'on Biol\'ogica de Do\~nana (EBD-CSIC)\\
  Seville, Spain
 }

\begin{document}
\maketitle
\baselineskip=8.5 mm

\newpage

{\bf The Web of Life (www.web-of-life.es) provides a graphical user interface, based on Google Maps, for easily visualizing and downloading data on ecological networks of species interactions. It is designed and implemented in a relational database management system, allowing sophisticated and user-friendly searching across networks. Users can access the database by any web browser using a variety of operating systems. Data can be downloaded in several common formats, and a data transmission webservice in JavaScript Object Notation is also provided.}

\newpage
\section{Introduction}
In nature, species do not live in isolation but form large networks of interdependences that are often depicted as a set of nodes (species) connected by links (species interactions). This entangled web of life is increasingly threated by several drivers of global change (Tylianakis et al., 2008, 2010), such as climate warming (e.g., Memmott {\em et al.}, 2007), habitat loss and fragmentation (e.g., Tylianakis {\em et al.}, 2007), and invasive species (e.g., Aizen {\em et al.}, 2008). In this network context, species interactions are a component of biodiversity as important as species themselves (Thompson, 2009). We know we are losing species interactions (Aizen {\em et al.}, 2012), which may drive ecological communities towards tipping points (Lever {\em et al.}, 2014). We need, hence, a network thinking to predict future community-wide scenarios (Tylianakis {\em et al.}, 2008).

Since the 1980s, data on who interacts-with-whom in ecological communities have been compiled, focused mainly on food webs and plant-animal mutualistic networks. Data, however, can be found mainly as appendices or supplementary material in the papers where the authors originally published their work. The Interaction Web DataBase (www.nceas.ucsb.edu/interactionweb/) created in 2003 and hosted by the National Center for Ecological Analysis and Synthesis (NCEAS) at the University of California (USA) is, to our knowledge, the only initiative with the aim of centralizing the available information on ecological networks. However, as far as we know, it is currently outdated. In addition, because it only provides datasheets, searching options are not implemented. This limits the potential use of the database as a working tool to tackle scientific questions across networks. In addition, if we aim to reach stackeholders and policy makers, an easy way of visualizing the network dataset compiled throughout the world is of paramount importance to capture their attention.  

\section{Features}
We provide the most comprenhensive dataset of plant-animal mutualistic networks to date. In contrast to antagonistic interactions in which one species obtains a benefit at expenses of the other, in plant-animal mutualistic interactions, like those between a plant and a pollinator or between a plant and a seed disperser, both species obtain mutual benefit. Food webs and other ecological interactions such as host-parasite, host-parasitoid, plant-ant, plant-epiphyte, plant-herbivore, and anemone-fish interactions, will be available soon.

\subsection{Location map}
The user interface is based on Google Maps. Over the map, colored circles indicate where the compiled networks are located. Colors depict the type of ecological interaction. The user can zoom in or out and drag the map.

\subsection{Network datasheet}
Some basic information about the network can be obtained when the mouse pointer is over one of the colored circles. By left clicking on it, a detailed information about the network is dynamically generated: number of species, number of interactions, network connectance, locality, geographical coordinates, original reference, and a unique network identifier. The network identifier is intended to be adopted by the community of researchers as a unique tag to identify a given network across future studies. The network of interactions is also displayed along the species names (when they are identified). Depending on the data compiled by the researchers, a matrix of ones and zeros (presence/absence of the interaction, respectively) or a matrix of natural numbers indicating the number of visits performed by a pollinator species on a plant species, is displayed. A java script graphical representation using the D3js library is available for a quick visualization of the network.

\subsection{Data filtering criteria}
Network selection can be filtered directly from the menu bar by selecting the type of interaction (up to now pollination or seed dispersal), type of data (binary: presence/absence of the interactions, or weighted: number of visits), number of species, and number of interactions. The list of networks selected as resulting from the filtering criteria applied is immediately ready for visualization or download. Searchers across networks by species names can be performed from the list of networks selected, which allows a meta-analysis never accomplished before.

\subsection{Data download}
Network data can be downloaded from the location map and from the filtering criteria. Species names can also be included. The following file formats are available: comma-separated values (.csv), Excel spreadsheet format (.xls), Pajek format (.net) that can be imported in Gephi as well for visualization, and as Java Script Object Notation (.json). Downloading a large dataset could take some time because a zip file containing each network as a single file, a file for the references, and a readme file, is dynamically generated. A log file tracking the history of changes for the downloaded network dataset (if exists) is also included in the zip file.

\section{Implementation}
 The Web of Life has been designed and implemented in an open-source relational database management system (MySQL). This allows sophisticated and user-friendly searching across networks. It also provides an easy way of incorporating new network data available in the future. In order to minimize spelling mistakes when introducing new data, we do not provide an online interface for data entry, so that users cannot enter and edit their data directly. Users can access the database through any web browser using a variety of operating systems.

\section{Conclusion}
Biodiversity is much more than a list of species. Interactions among species are the {\em glue} of biodiversity. If we aim at predicting future community-wide scenarios and anticipating planetary critical transitions, we have to consider the entangled web of interactions among species. Here, we introduce The Web of Life, a database for visualizing and downloading data of species interaction networks. This repository allows scientists to do research within and between networks compiled at different places all over the world.

\section{Acknowledgements}
We would like to thank Jordi Bascompte's lab members for their suggestions during the design of the web, as well as the authors of the compiled networks for their invaluable effort during the fieldwork. This web service is supported by the European Research Council under the European Community's Seventh Framework Programme (FP7/2007-2013) through an Advanced Grant to Jordi Bascompte (grant agreement 268543). M.A.F. holds a postdoctoral fellowship (JAE-Doc) from the Program ``Junta para la Ampliacion de Estudios'' co-funded by the Fondo Social Europeo (FSE).

\newpage
\section{References}
\noindent Aizen, M. A., Morales, C., and Morales, J. (2008). Invasive mutualists erode native pollination webs. {\em PLoS Biology}, 6:e31.

\noindent Aizen, M. A., Sabatino, M., and Tylianakis, J. M. (2012). Specialization and rarity predict nonrandom loss of interactions from mutualist networks. {\em Science}, 335:1486-1489.

\noindent Lever, J., van Nes, E. H., Scheffer, M., and Bascompte, J. (2014). The sudden collapse of pollinator communities. {\em Ecol. Lett.}, 17:350-359.

\noindent Memmott, J., Craze, P. G., Waser, N. M., and Price, M. V. (2007). Global warming and the disruption of plant-pollinator interactions. {\em Ecol. Lett.}, 10:710-717.

\noindent Thompson, J. N. (2009). The coevolving web of life. {\em Am. Nat.}, 173:125-140. 

\noindent Tylianakis, J. M., Tscharntke, T., and Lewis, O. T. (2007). Habitat modification alters the structure of tropical host-parasitoid food webs. {\em Nature}, 445:202-205.

\noindent Tylianakis, J. M., Didham, R. K., Bascompte, J., and Wardle, D. A. (2008). Global change and species interactions in terrestrial ecosystems. {\em Ecol. Lett.}, 11:1351-1363.

\noindent Tylianakis, J. M., Lalibert\'e, E., Nielsen, A., and Bascompte, J. (2010). Conservation of species interaction networks. {\em Biol. Cons.}, 143:2270-2279.

\newpage
\section{Figures}
\begin{figure}[!ht]
  \includegraphics[width=1.00\textwidth]{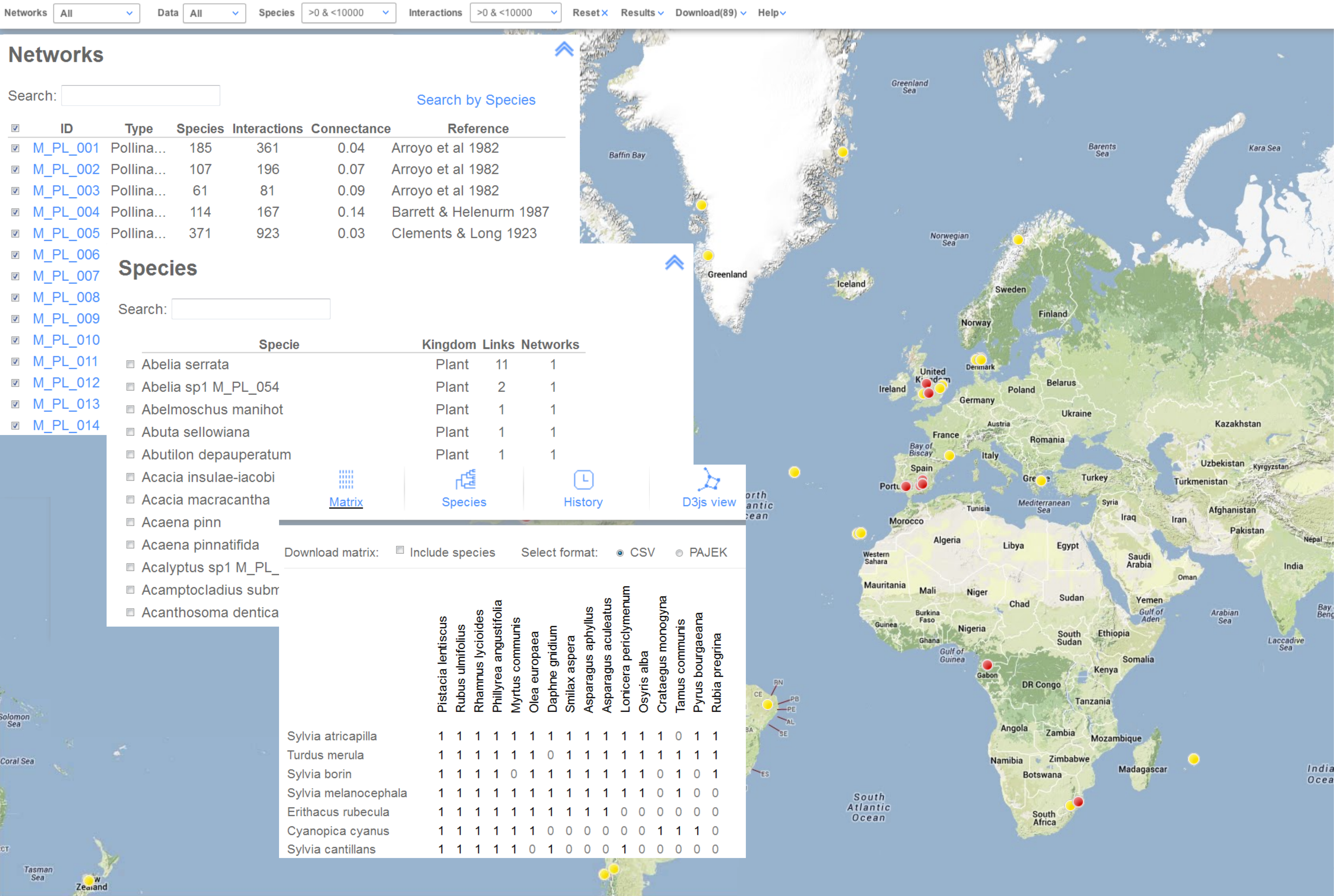}
  \caption{Snapshot of The Web of Life. On top, the menu bar showing the data filtering criteria. Over the map, three pop-up windows are displayed: the network list containing the selected networks, the species list for searching across networks, and an example of a network datasheet showing the presence/absence of mutualistic interactions between plant and animal species (from back to front, respectively).}
\end{figure}




\end{document}